\documentclass[preprint]{elsarticle}
\usepackage{amsmath,hyperref,color,listings}

\makeatletter
\def\ps@pprintTitle{%
 \let\@oddhead\@empty
 \let\@evenhead\@empty
 \def\@oddfoot{}%
 \let\@evenfoot\@oddfoot}
\makeatother

\def\Li{\text{Li}}
\def\Litt{\text{Li}_{22}}
\def\ep{\epsilon}
\def\lisk{\texttt{LiSK} }

\begin{document}
	
	\begin{frontmatter}
		
		\title{\vspace{-2cm}\hfill {\small\rm TTK-16-18}\\ \vspace{1cm} LiSK - A \texttt{C++} Library for Evaluating Classical Polylogarithms and $\Litt$}
		\author{Sebastian Kirchner}
		\address{Institut f\"ur Theoretische Teilchenphysik und Kosmologie\\ RWTH Aachen University\\Aachen, Germany}
		\ead{kirchner@physik.rwth-aachen.de}
		\ead[url]{https://bitbucket.org/SebastianKirchner/lisk}
		
		\begin{abstract}
		I present a lightweight \texttt{C++} library for the evaluation of classical polylogarithms $\Li_n$ and the special function $\Litt$ for arbitrary complex arguments. The evaluation is possible in arbitrary precision arithmetic and features also an explicit \texttt{double} precision implementation for a much faster numerical evaluation. The implementation is based on Ref~\cite{Frellesvig:2016ske}.
		\end{abstract}
		
		\begin{keyword}
		Classical polylogarithms, $\Litt$, \texttt{C++}
		\end{keyword}
		
	\end{frontmatter}
	
\section{Introduction} 
\label{sec:introduction}

Classical polylogarithms and their multi-variable generalisations to generalised polylogarithms (GPLs) play an important role in modern quantum field theory. These functions and specialisations thereof have been studied in early works by Poincar\'e and Kummer, much later implicitly by Chen in his works on iterated integrals~\cite{Chen:1977oja} and by Goncharov~\cite{Goncharov:1998kja,Goncharov:2001iea,Goncharov:2010jf}. The importance lies in the fact that a large class of dimensionally regulated loop Feynman integrals in $d=4-2\ep$ dimensions have been found to be expressible in terms of these GPLs (Please refer to Refs~\cite{Duhr:2014woa,Henn:2014loa,Henn:2014qga} and references therein for more detailed discussions). It is known that all one-loop integrals up to $\ep^0$ can be written in terms of the logarithm and the dilogarithm $\Li_2$~\cite{Ellis:2007qk}. The situation at higher orders in the $\ep$ expansion or, similar, at higher orders in perturbation theory, is already more involved. However, it was in Ref~\cite{Frellesvig:2016ske} recently proven that all GPLs up to weight four can be expressed in terms of classical polylogarithms and the special function $\Litt$. Hence the evaluation of Feynman integrals, fulfilling the upper restrictions, and GPLs, in general, reduce then to two fundamental steps. The first is expressing all GPLs up to weight four by $\Li_n$ with $n\le 4$ and $\Litt$. Although this step is far from trivial because divergent intermediate expressions may occur which have to be suitably regulated, a handful of methods have been proposed which, in principle, allow for a complete automatisation of this step. See e.g. Refs~\cite{Frellesvig:2016ske,Goncharov:2010jf,Duhr:2014woa,Goncharov.A.B.:2009tja,Duhr:2012fh,Bogner:2016qbf,Bogner:2012dn}. The second step is the fast and numerically stable evaluation of the resulting $\Li_n$ and $\Litt$. I want to address the latter in this work.

Efficient algorithms for the numerical evaluation of classical polylogarithms and specific subclasses of GPLs have been developed and studied in the literature~\cite{Frellesvig:2016ske,Kolbig:1969zza,'tHooft:1978xw,Borwein:1999js,Gehrmann:2001pz,Gehrmann:2001jv}. The evaluation of GPLs, and $\Litt$ for that matter, with arbitrary complex arguments, however, is only publicly available in \texttt{GiNaC}~\cite{Bauer:2000cp,Vollinga:2004sn}. That specific implementation allows to evaluate GPLs without restrictions on the weight in arbitrary precision via the \texttt{CLN} library. Using the arbitrary precision implementation without exceptions obviously yields sincere performance penalties if the user is solely interested in standard floating point formats as e.g. \texttt{double} precision. The authors of Ref.~\cite{Frellesvig:2016ske} provide a new algorithm and also some \texttt{C++} routines for the evaluation of the classical polylogarithms and the special function $\Litt$. However, their implementation is limited to weights $n\le 6$ of the $\Li_n$ and is exclusively given in \texttt{double} precision. Experience shows that higher precision results are often required in the computation of multi-loop Feynman integrals, such as in Ref~\cite{Gehrmann:2015ora,vonManteuffel:2015msa}. 

I believe that a combination of both worlds is desirable. The \texttt{template} features of the modern \texttt{C++} language allows to easily incorporate multiple supported types without duplicating the entire code. The current version of \lisk supports a very fast \texttt{double} precision implementation as well as an arbitrary precision implementation utilising the \texttt{CLN} library. \lisk implements the computation of $\Li_n$ and $\Litt$ for arbitrary complex arguments using the algorithms from Ref~\cite{Frellesvig:2016ske}; but extends the computation of $\Li_n$ to arbitrary weights.

\lisk has been tested under \texttt{SUSE Linux} and is known to work with the compiler versions \texttt{g++-4.8.3}, \texttt{clang++-3.5} and \texttt{icc 14.0.2} using the GNU C++ standard library. \lisk can be downloaded from
\begin{center}
	\url{https://bitbucket.org/SebastianKirchner/lisk}
\end{center}
	
This work is structured as follows. I will explain the user interface and show an example how to use \lisk in Sec.~\ref{sec:usage}. In Sec.~\ref{sec:checks} I will validate the results obtained by \lisk and conclude in Sec.~\ref{sec:conclusions}.
	

\section{Usage} 
\label{sec:usage}

\lstset{language=C,frame=single,basicstyle=\small}

\lisk is a complete \emph{header-only} library, i.e. the library does not have to be built separately but only the header file must be included in the user's code via the usual
\begin{lstlisting}
#include "lisk.hpp"
\end{lstlisting}
\underline{Please note}: The implementation has been put into the \emph{lisk.cpp} file due to readability. This source file must be present in the same directory as the header file.

\lisk is completely encapsulated in the namespace \texttt{LiSK} and relies on several \texttt{C++-11} features. A central part is the \textbf{CLN}\footnote{\texttt{CLN} can be downloaded from \url{http://www.ginac.de/CLN/}} library by Bruno Haible and has to be linked to all programs using \texttt{LiSK}; also if using only \texttt{double} precision.
A simple example of how to use \lisk is given in \texttt{example/example.cpp}, which can be compiled via
\begin{lstlisting}
g++ -O2 -std=c++11 example.cpp -o example -I/cln/include/path 
-I../LiSK -L/cln/library/path  -lcln
\end{lstlisting}
Alternatively, if \texttt{cmake} is available, simply by running
\begin{lstlisting}
cmake .
\end{lstlisting}
in the \texttt{example} directory. If \texttt{CLN} is not found automatically the \emph{CLN\_INCLUDE\_DIR} and \emph{CLN\_LIB} path can easily be set via
\begin{lstlisting}
ccmake .
\end{lstlisting}

I will illustrate the usage of \lisk on this simple example. In the following \texttt{T} denotes one of the two currently supported types \texttt{std::complex<double>} and \texttt{cln::cl\_N}. The first action should be to create a \lisk object of type \texttt{T} and precision \texttt{p} via
\begin{lstlisting}
LiSK::LiSK<T> lisk(n,p);
\end{lstlisting}

During object creation \lisk is initialised and all constants required for the computation of the $\Li_n(x)$ and $\Litt(x,y)$ are pre-computed. This reflects the main idea of \texttt{LiSK}; prepare and save all needed constants during its initialisation phase and use them during the actual computation. The constructor  \texttt{LiSK$(\text{n}=4,\text{prec}=34)$} features two optional arguments.

The first argument \emph{n} defines the weight of the $\Li_n(x)$ up to which the constants are computed during \texttt{LiSK}s initialisation phase. It is not mandatory but advised to set \emph{n} to a value which resembles the highest expected weights. If higher weights are encountered during the computation the constants will be adapted dynamically. This, obviously, leads to longer evaluation times for the polylogarithms for which the higher weights have been encountered. This situation should be avoided as much as possible.

The second argument of the constructor sets the desired precision if $\texttt{T}=\texttt{cln::cl\_N}$ is chosen. This argument is superfluous in the double precision case. E.g. set $p=34$ to obtain results with 34 digit precision. Internally all floating point values are set to this precision. This is also true for the initial complex arguments x and y supplied by the user. The user has to ensure that the supplied input values match the requested precision.

Calling the public wrapper functions for the computation of $\Li_m(x)$ for positive integer weights \emph{m} and $\Litt(x,y)$ at given points x and y is given by
\begin{lstlisting}
lisk.Li(m,x);
lisk.Li22(x,y);	
\end{lstlisting}
where \emph{m} should be smaller than \emph{n}. One might also call 
\begin{lstlisting}
LiSK::LiSK<T>(n,p).Li(m,x);
LiSK::LiSK<T>(n,p).Li22(x,y);	
\end{lstlisting}
but this is strongly \underline{not} recommended due to the above mentioned reasons. However, special wrapper functions exist for the classical polylogarithms with weights $n \leq 4$
\begin{lstlisting}
lisk.Li1(x);
lisk.Li2(x);
lisk.Li3(x);
lisk.Li4(x);
\end{lstlisting}

In case some error is encountered \lisk will throw a \texttt{std::runtime\_error}. Hence, it is advised to put all calls to \lisk into a \texttt{try}-block like
\begin{lstlisting}
try{
	/* some code */
}
catch(std::runtime_error &e){
	std::cout << e.what() << std::endl;
}	
\end{lstlisting}

Last but not least it must be ensured that all expressions, initial and intermediate, are well defined. To this end a small positive imaginary part is added to the initial arguments $x$ and $y$ of $\Li_n(x)$ and $\Litt(x,y)$, i.e. $x\to x - i\ep$. The sign of the imaginary part is chosen according to the widely used convention for mathematical software\footnote{"$[\cdots]$ implementations shall map a cut so the function is continuous as the cut is approached coming around the finite endpoint of the cut in a counter clockwise direction"~\cite{british2003c}}, which agrees for example with the convention used in \texttt{Mathematica} and \texttt{GiNaC}.
The value of $\ep$ is set to $10^{-(p-\text{\emph{\_offset}})}$ during initialisation. Hereby defines $p$ the requested precision in the constructor ($p=17$ in \texttt{double} precision mode). The default value of \emph{\_offset} is $2$. The user can change this value at the top of the \lisk header\footnote{A handful of options to change the behaviour of \lisk can be set in the header. However, these are of rather technical (and experimental) nature and will not be discussed here. The user is advised to proceed with caution.}.


\section{Checks} 
\label{sec:checks}

In this section I discuss some checks of the \lisk library by its comparison against \texttt{GiNaC}. 

Various tests in different parameter regions have been performed. Similar to Ref~\cite{Frellesvig:2016ske} I have checked the level of precision by evaluating $\Li_n$ (for multiple values of $n$) and $\Litt$ at $10^4$ random parameter points, where the absolute value of each point is exponentially distributed between $10^{-10}$ and $10^{10}$. Multiple tests for real input parameters and parameters in vicinity of zero and one have been performed. In case of the \texttt{double} precision implementation all points had a relative deviation ($2 |a-b|/|a+b|$) smaller than $10^{-13}$. The arbitrary precision implementation is in perfect agreement with \texttt{GiNaC}, provided that a consistent $i\ep$ prescription is given in both programs.

From the performance point of view \lisk evaluates the $\Li_n$ and $\Litt$ in comparable time and orders of magnitudes faster than the current \texttt{GiNaC} implementation; depending on the parameter values. The exact timing depends on the used architecture, compiler, program setup, etc. For brevity of this work I encourage the interested user to try and compare both implementations for his specific problem. The \texttt{double} precision implementation is, as expected, much faster than using the arbitrary precision arithmetic in \texttt{CLN} and might lead to a sizeable speed-up in computations where \texttt{double} precision is sufficient.


\section{Conclusions} 
\label{sec:conclusions}

In this work I have presented the first version of \texttt{LiSK}, an independent implementation for the numerical evaluation of $\Li_n$ and $\Litt$ at arbitrary complex arguments. \lisk evaluates $\Li_n$ and $\Litt$ in arbitrary precision but also features an explicit \texttt{double} precision implementation. The latter provides a much faster evaluation of the polylogarithms, which might be desirable for certain users. The complete \texttt{template} \texttt{C++} implementation provides a basis to easily extend the supported types, e.g. to the GNU Quad-Precision Math Library.

\lisk utilises many of the algorithms from Ref~\cite{Frellesvig:2016ske}. However not all of these algorithms are currently in use. Especially the numerical evaluation of $\Litt$ with arguments in vicinity of one leaves room for performance improvements in upcoming versions.


\section{Acknowledgements} 
\label{sec:acknowledgements}

This work was supported by the Deutsche Forschungsgemeinschaft through Graduiertenkolleg GRK 1675.

	

\bibliography{manual.bib}

	
\end{document}